\documentclass[twocolumn,showpacs,preprintnumbers,amsmath,amssymb]{revtex4}
\usepackage{amsmath}
\usepackage{amssymb}
\usepackage{amsfonts}
\usepackage{graphics}
\usepackage{dcolumn}% Align table columns on decimal point
\usepackage{bm}% bold math
\def\beq{\begin{equation}}
\def\eeq{\end{equation}}
\def\beqarray{\begin{eqnarray}}
\def\eeqarray{\end{eqnarray}}

\begin{document}

\title{Wavelet Methods in the Relativistic Three-Body Problem}

\author{Fatih Bulut}
\altaffiliation[Also at ]{Department of Physics and Astronomy, The 
University of Iowa.}%Lines break automatically or can be forced with \\
\author{W. N. Polyzou}

\email{polyzou@uiowa.edu}
\affiliation{%
Department of Physics and Astronomy, \\The 
University of Iowa   
}%

\date{\today}

\begin{abstract}

Abstract: In this paper we discuss the use of wavelet bases to solve
the relativistic three-body problem.  Wavelet bases can be used to
transform momentum-space scattering integral equations into an
approximate system of linear equations with a sparse matrix.  This has
the potential to reduce the size of realistic three-body calculations
with minimal loss of accuracy.  The wavelet method leads to a clean,
interaction independent treatment of the scattering singularities
which does not require any subtractions.

\end {abstract}

\pacs{03.65Pm,11.30.Cp,11.80-m,21.45.+v,24.10Jv,25.10.+s}
\keywords{wavelets,relativistic three-body models}%

\maketitle
\section{Introduction}

This is the third paper \cite{kessler03}\cite{kessler04} in a series
of investigations designed to explore potential advantages of using
wavelet numerical analysis to solve the relativistic three-body
problem.  Commercially, wavelets are used to convert raw digitized
photographic images to compressed JPEG files \cite{jpeg}.  In this
application the data compression leads to a large saving in storage
space with a minimal loss of information.  The compression involves
expanding the raw digital image in a wavelet basis and setting the
smaller expansion coefficients to zero.  The kernel of a scattering
integral equation and a raw digital image can both be approximated by
rectangular arrays of numbers with some continuity properties.  This
suggests that the bases used to compress digital images could be used
to generate accurate sparse matrix approximations to the kernel.

The ability to construct numerically exact solutions to the quantum
mechanical three-body problem coupled with the ability to accurately
measure complete sets of experimental observables constrains the form
of the three-nucleon Hamiltonian.  These constraints have resulted in
the construction of realistic model nucleon-nucleon interactions
\cite{deSwart}\cite{v18} \cite{cdbonn}.  When these interactions are
used in the many-nucleon Hamiltonian, the resulting dynamical model
provides a good quantitative description of low-energy nuclear physics
\cite{lightnuc}. 

The state of the art in few-body computations has improved to the
point where numerically exact scattering calculations at energy and
momentum transfers of hundreds of $MeV$ have been performed
\cite{walter}.  Higher energy calculations are possible.  As in the
low-energy case, the structure of Hamiltonians for higher energy
reactions can be constrained by the consistency of the few-body
calculations with precise measurements of complete sets of
experimental observables.

The success of the few-body approach to low-energy nuclear physics is
a consequence of (1) knowing the relevant degrees of freedom
(nucleons), (2) working with the most general Hamiltonians involving
these degrees of freedom that are consistent with the symmetries of
the system (Galilean invariance) and (3) understanding the relation
between the few and many-body problem (cluster properties).  To extend
this success to reactions involving higher energy scales (1) the
relevant degrees of freedom may have to include explicit mesonic or
sub-nucleonic degrees of freedom (2) Galilean invariance must be
replaced by Poincar\'e invariance and (3) cluster properties must be
maintained.

Each of the required extensions of low-energy nuclear dynamics is
non-trivial, and progress has been made on all three problems
\cite{walter}\cite{wp91}\cite{wp02}\cite{fuda}\cite{wp03}.  The
purpose of this paper is to focus on technical aspects of using
wavelet numerical analysis to construct exact numerical solutions of
the dynamics in Poincar\'e invariant few-body models.  While the scope
of this paper is limited to three-nucleon models with no explicit
mesonic or subnucleonic degrees of freedom and S-matrix cluster
properties \cite{fritz65}, the formulation and advantage of methods
discussed in this paper are straightforward to extend to systems with
explicit mesonic degrees of freedom and stronger forms of cluster
properties \cite{wp03}.  Moving singularities are a generic feature of
the dynamical equations in all of these cases.

Relativistic few-body equations are naturally formulated in momentum
space.  Relativistic kinematic factors, Wigner rotations and Melosh
rotations are all multiplication operators in momentum space.  The
compactness of the iterated Faddeev-Lovelace kernel implies the kernel
of the integral equations can be uniformly approximated by a finite
matrix, resulting in a finite linear system.  In the momentum
representation these linear systems have large dense matrices, which
increase in size with increasing energy and momentum transfer.  It is
desirable to be able to perform accurate calculations at energy and
momentum scales where subnuclear degrees of freedom are relevant.  At
these scales a relativistic treatment is required and advances in
computational efficiency are needed to perform realistic calculations.
The ability of the wavelet transform to efficiently transform a
dense matrix to an approximate sparse matrix suggests that wavelet methods
can provide a powerful tool for improving the efficiency of relativistic 
few-body computations.

The advantages of using wavelet numerical analysis to solve momentum
space scattering integral equations was investigated in
\cite{kessler03} and \cite{kessler04}.  These papers used wavelet
numerical analysis to solve the Lippmann-Schwinger equation for a
system of two nucleons interacting with a Malfliet Tjon V potential
\cite{Tjon}\cite{Payne} using partial wave expansions \cite{kessler03}
and direct integration \cite{kessler04}.  In both applications the
kernel of the integral equation was accurately approximated by a
sparse matrix, which resulted in accurate approximate solutions.  The
success of these applications indicates that wavelet numerical
analysis will have similar advantages when applied to the relativistic
three-body problem.

The feature of the three-body problem that is not present in the
two-body applications is moving singularities.  The methods used in
\cite{kessler03} and \cite{kessler04} are not applicable to problems
with moving singularities.  The purpose of this paper is to illustrate
how to apply wavelet numerical analysis to treat the moving
singularities that appear in the relativistic three-body problem.

\section{Overview - Wavelet Numerical Analysis}

The applications in ref. \cite{kessler03} and \cite{kessler04} used
Daubechies' wavelets.  These wavelets differ from the wavelets
used to store JPEG images.  The Daubechies' wavelets are orthonormal
but the basis functions are not reflection symmetric; while the
wavelets used to store JPEG images sacrifice orthonormality to obtain
more symmetric basis functions.  The results of
ref. \cite{kessler03} and \cite{kessler04} indicate that the
Daubechies' wavelets are suitable for scattering calculations.

Daubechies' wavelets \cite{Daubechies}\cite{Daubechies2} are discussed
in many texts on wavelets \cite{Kaiser} \cite{Resnik} \cite{Strang}
\cite{jorgensen}.  They are fractal functions that have complex
structures on all scales.  Because the basis functions have structure
on all scales, numerical applications with wavelets require a
different approach to numerical analysis, hence the term wavelet
numerical analysis.

We use the Daubechies' wavelets because they are a dense orthonormal
set of compactly supported functions with the property that finite
linear combinations can locally pointwise represent low-degree
polynomials.

Two types of functions are needed to generate wavelet bases.  These
functions are called scaling functions and wavelets.  The
scaling function, $\phi (x)$, is the solution of the linear
renormalization group equation:
\begin{equation}
D \phi (x) = \sum_{l=0}^{2K-1} h_l T^l \phi (x) 
\label{eq:CA}
\end{equation}
with normalization 
\beq
\int_{- \infty}^{\infty}  \phi (x) dx =1 . 
\label{eq:CB}
\eeq
Equation (\ref{eq:CA}) is called the scaling equation.    

In equation ({\ref{eq:CA}) $D$ is the unitary scaling operator
\beq
Df(x) := {1 \over \sqrt{2}} f ({x \over 2}) 
\label{eq:AC}
\eeq
which stretches the support of the function by a factor of two.
The operator $T$ is the unitary unit translation operator
\beq
T f(x) = f(x-1) .
\label{eq:CD}
\eeq 
The coefficients $h_l$ are real numbers that determine the properties
of the scaling function.  $K$ is a finite positive integer.  The
calculations in \cite{kessler03}\cite{kessler04} used Daubechies' $K=3$
wavelets.  The reason for this choice will be discussed later.
For the $K=3$ Daubechies' wavelets the six scaling coefficients $h_l$ are 
given in Table 1.

\begin{center}
\begin{table} % [hbt]
{\bf Table 1: Daubechies' $K=3$ Scaling Coefficients } \\[1.0ex]
\begin{tabular}{c c}
\hline
\hline
$h_l$ & K=3   \\
\hline					      		      
$h_0$ &$(1+\sqrt{10}+\sqrt{5+2\sqrt{10}})/16\sqrt{2}$ \\
$h_1$ & $(5+\sqrt{10}+3\sqrt{5+2\sqrt{10}})/16\sqrt{2}$ \\
$h_2$ & $(10-2\sqrt{10}+2\sqrt{5+2\sqrt{10}})/16\sqrt{2}$ \\
$h_3$ & $ (10-2\sqrt{10}-2\sqrt{5+2\sqrt{10}})/16\sqrt{2} $ \\
$h_4$ & $(5+\sqrt{10}-3\sqrt{5+2\sqrt{10}})/16\sqrt{2}$ \\
$h_5$ & $(1+\sqrt{10}-\sqrt{5+2\sqrt{10}})/16\sqrt{2}$ \\
\hline
\hline
\end{tabular}
% \label{tabspec}
\end{table}
\end{center}

The fractal structure of $\phi (x)$ is a consequence of the 
scaling equation (\ref{eq:CA})
which shows that the scaling function on a given scale is a finite
linear combination of translates of the same function on half the
scale.  

The scaling equation implies that the scaling coefficients 
$h_l$ satisfy 
\beq
\sum_{l=0}^{2K-1} h_l  = \sqrt{2}
\label{eq:CE}
\eeq
and that the solution $\phi (x)$ of equation (\ref{eq:CA}) 
has support on the interval $[0,2K-1]$ \cite{waverev}.

The unit translates of the scaling function 
are orthonormal 
\beq
(T^m\phi , T^n \phi ) = \delta_{mn} 
\label{eq:CF}
\eeq
provided the scaling coefficients satisfy the 
additional constraints:
\beq
\sum_{l=0}^{2K-1} h_l h_{l-2m}  = \delta_{m0}.
\label{eq:CG}
\eeq

The scaling function $\phi (x)$ is continuous (for $K>1$) and can be
computed exactly at all dyadic rationals using equations (\ref{eq:CA})
and (\ref{eq:CB}).  This method is used to compute the Daubechies' 
$K=3$ scaling function plotted in Figure 1.

\begin{figure}
\begin{center}
\rotatebox{270}{\resizebox{2.9in}{!}{
\includegraphics{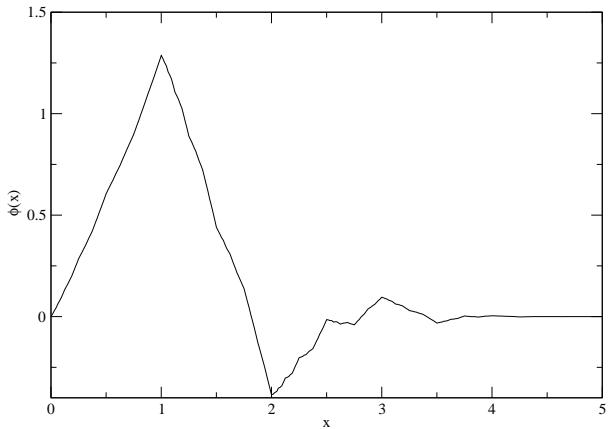}}
}
\end{center}
\label{Fig:1}
\caption{Daubechies' $K=3$ scaling function.}
\end{figure}

The subspace of square integrable functions on the real line that 
can be expressed as linear combinations integer translates of the 
scaling function, $T^n\phi (x)$, is the subspace ${\cal V}_0$ 
of $L^2 ({\mathbb R})$ defined by :
\beq
{\cal V}_0 := \{ f(x) = \sum_{n=-\infty}^{\infty} f_n T^n \phi (x) 
\vert \sum_{n=-\infty}^{\infty} \vert f_n \vert^2 < \infty\}.
\label{eq:CH}
\eeq

Application of powers of the scaling operator 
$D^k$ to ${\cal V}_0$ defines subspaces ${\cal V}_k$ 
with coarser $(k>0)$ or finer resolution $(k< 0)$:
\beq
{\cal V}_k = D^k {\cal V}_0 .
\label{eq:CI}
\eeq
The space ${\cal V}_k$ is called the approximation space with resolution 
$k$.  The resolution determines the size of the smallest features that
can be approximated by functions in ${\cal V}_k$.

The scaling functions 
\beq
\phi_{kn} (x) := D^k T^n \phi (x) = 
{1 \over 2^{k/2}}\phi ( {x \over 2^k}-n)
\label{eq:CJ}
\eeq
are an orthonormal basis for ${\cal V}_k$.  The support of $\phi_{kn}(x)$ is 
$[2^kn , 2^k(n+2K-1)]$. 
       
The scaling equation implies the inclusions
\beq
{\cal V}_k \supset {\cal V}_{k+1}.  
\label{eq:CK}
\eeq 
The orthogonal compliment of ${\cal V}_{k+1}$ in ${\cal V}_{k}$
is denoted by ${\cal W}_{k+1}$ which leads to the orthogonal 
decomposition
\beq
{\cal V}_k = {\cal V}_{k+1} \oplus {\cal W}_{k+1}.   
\label{eq:CL}
\eeq 

Orthonormal basis functions, $\psi_{km}(x)$, for the subspaces 
${\cal W}_k$ are elements of ${\cal V}_{k-1}$ given by
\beq
\psi_{km} (x) = D^k T^m \psi (x) 
\eeq
\beq
\psi (x) =  \sum_{l=0}^{2K-1} g_l D^{-1} T^{l} \phi(x)   
\label{eq:CM}
\eeq
where 
\beq
g_l = (-)^l h_{2K-1 -l} .
\label{eq:CN}
\eeq
The subspaces ${\cal W}_{kn}$ are called the wavelet spaces and the
basis functions $\psi_{kn}$ are called wavelets.  The support of
$\psi_{kn}(x)$ is identical to the support of $\phi_{kn}(x)$.  Since
the wavelets are finite linear combinations scaling functions they are
also fractal functions.

The function $\psi(x)=\psi_{00}(x)$ is called the mother wavelet; 
the Daubechies' $K=3$ mother wavelet is shown in Figure 2.

\begin{figure}
\begin{center}
\rotatebox{270}{\resizebox{2.9in}{!}{
\includegraphics{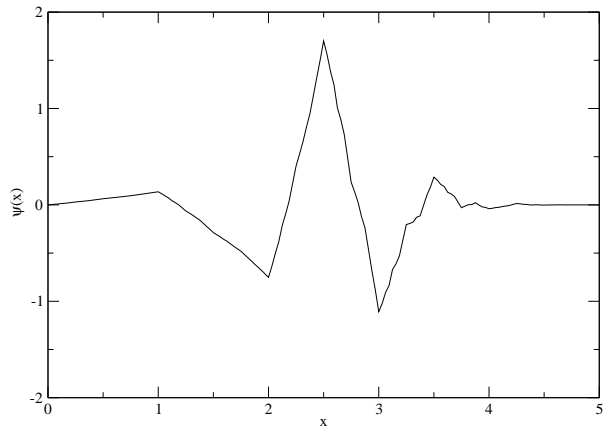}}
}
\end{center}
\label{Fig:2}
\caption{Daubechies' $K=3$ mother wavelet.}
\end{figure}

The coefficients $h_l$ for the Daubechies' $K$ wavelets are determined
by equations (\ref{eq:CE}) and (\ref{eq:CG}) and the requirements
\beq
\int_{-\infty}^{\infty} 
\psi (x) x^n dx =0; \qquad n=0,1,\cdots, K-1 
\label{eq:CO} 
\eeq
which implies that $\psi_{km} (x)$ is locally orthogonal to all 
polynomials of degree $K-1$.  Equation (\ref{eq:CO}) 
can be expressed directly in terms of the scaling coefficients:
\[
\sum_{l=0}^{2K-1}  l^k g_l =  
\]
\beq
\sum_{l=0}^{2K-1}  l^k (-)^l h_{2K-1-l} =  0 
\qquad k=0 \cdots K-1.
\label{eq:COA}
\eeq
For any $K>0$  conditions (\ref{eq:CE}), (\ref{eq:CG}) and 
(\ref{eq:COA}) determine $h_l$  
up to reflection,
\beq
h_l \to h'_l := h_{2K-1-l} .
\label{eq:CP}
\eeq
The entries in Table 1 are the solution of these equations for $K=3$.
The Daubechies' wavelets have the property that as $k\to - \infty$ 
(infinitely fine resolution) the
space ${\cal V}_k$ becomes all of $L^2 (\mathbb{R})$.

The decomposition (\ref{eq:CL}) implies that 
\beq
{\cal V}_k = {\cal W}_{k+1} \oplus {\cal W}_{k+2} \cdots
{\cal W}_{k+m-1} \oplus  {\cal W}_{k+m}\oplus {\cal V}_{k+m},  
\label{eq:CQ}
\eeq
for any $m>0$.  This means that functions in the approximation space
${\cal V}_k$ can be expanded as linear combinations of the scaling
basis functions of resolution $k$ or equivalently as linear
combinations of the scaling basis functions of a coarser resolution
$k'=k+m$ and wavelet basis functions of all resolutions from $k+1$ to
$k+m$.

If we let $k\to - \infty$ with $m+k$  finite, then the functions in
(\ref{eq:CQ}) become a basis for $L^2({\mathbb R})$.  Since the
$\psi_{kn}(x)$ basis functions are locally orthogonal to 
degree $K-1$ polynomials, and
only a finite number of the $\phi_{kn}(x)$ are non-zero at any $x$,
it follows that finite linear combinations of $\phi_{k,n}(x)$
must be able to locally pointwise represent degree $K-1$ polynomials.
%Furthermore, since only a finite number of the
%$\phi_{k,n} (x)$ are non-zero at any point, the local expansions are
%finite.

Thus, for the Daubechies' $K=3$-wavelets finite linear combinations of
the scaling basis functions $\phi_{kn}(x)$ can locally pointwise
represent polynomials of degree 2, while
the wavelet basis functions $\psi_{kn}(x)$ are orthogonal to
degree 2 polynomials.

Equation (\ref{eq:CQ}) implies that the projection $P_k$ of a function
$f(x)$ on ${\cal V}_k$ can be represented by
\beq
P_k f (x) = \sum_n a_n \phi_{kn} (x)  \qquad 
a_n = \int f(x) \phi_{kn}(x) dx 
\label{eq:CR}
\eeq
or equivalently
\beq
P_k f (x) = \sum_n a_n \phi_{k+m,n} (x)
+ \sum_{k'=k+1}^{k+m} b_{k'n} \psi_{k',n},
\label{eq:CS}
\eeq
\beq  
b_{kn} = \int f(x) \psi_{k,n}(x) dx  .
\label{eq:CT}
\eeq

For a sufficiently fine resolution (large $-k$) the scaling basis
functions have small support and integrate to a constant.  If $f(x)$
varies slowly on intervals of width $(2K-1)2^k$ then expansion
coefficients $a_n$ are well approximated by evaluating $f(x)$ at any
point in the support of $\phi_{kn} (x)$ and multiplying by $2^{k/2}$,
which is the integral of $\phi_{kn}(x)$.  This means that the scaling
function basis coefficients, $a_n$, are well approximated, up to a
fixed multiplicative constant, by sampling the original function.
These coefficients play the role of the raw image in a digital
photograph.  They provide an accurate, but inefficient approximation
of the function $f(x)$.

In the representation (\ref{eq:CS}), if $f(x)$ can be accurately
approximated by a polynomial of degree $K-1$ on the support of
$\psi_{kn}(x)$ then $b_{kn} \approx 0$.  This means that if $f(x)$ can
be well approximated by low-degree local polynomials with compact support on
multiple scales, most of the coefficients $b_{kn}$ will be be small
and the function can be accurately approximated by replacing these
small coefficients by zero.  The mean square error is the sum of the
squares of the discarded coefficients, which can be controlled by
selecting a maximum size of the discarded coefficients.  Even though
most of the basis functions in the representation (\ref{eq:CS}) are
orthogonal to low degree polynomials, the equivalence between the
representations (\ref{eq:CR}) and (\ref{eq:CS}) means that the
representation (\ref{eq:CS}) can still locally pointwise represent low
degree polynomials.  The orthogonal transformation connecting
equivalent representations (\ref{eq:CR}) and (\ref{eq:CS}) of ${\cal
V}_k$ is called the wavelet transform \cite{numrec}.  For $N$ basis
functions it can be implemented in $O(N)$ steps, which for large $N$
requires less steps than a fast Fourier transform. 

The scaling equation and normalization condition can be used to derive
exact expressions for the moments and partial moments of the scaling 
function
\beq
\langle x_{kn}^m \rangle := \int \phi_{kn} (x) x^m  dx
\label{eq:CU}
\eeq
\beq
\langle x_{kn}^m := \rangle_{[l,l']}\int_{2^k l}^{2^k l'}\phi_{kn} (x) x^m  
\qquad n\leq l,l' \leq 2K-1+n
\label{eq:CV}
\eeq
in terms of the scaling coefficients $h_l$.  Explicit expressions for the 
moments and partial moments appear in \cite{kessler03}, 
\cite{kessler04}, \cite{waverev}.

For the Daubechies' $K$ wavelets with $K>1$ the second moment of the
scaling function is the square of the first moment.  This means that
for $K=3$ the first moment provides a single quadrature point that will
integrate the scaling function times any second degree polynomial
exactly:
\beq
\int \phi (x) (a+bx+cx^2)  dx = a + b\langle x \rangle +c
\langle x \rangle^2 .  
\label{eq:CW}
\eeq
This is called the one-point quadrature.  Translating and rescaling
leads to one-point quadratures for all of the scaling basis functions
$\phi_{kn}(x)$.  The choice of $K=3$ Daubechies' wavelets in
\cite{kessler03} and \cite{kessler04} is motivated by their ability to
locally pointwise represent second degree polynomials and to exactly
integrate these local polynomials with a one-point quadrature.

In \cite{kessler03}, \cite{kessler04} and \cite{waverev}
the moments
and the scaling equation are used to compute the singular integrals 
\beq
L_{kn}^{\pm}  := \int dx {\phi_{kn}(x) \over x \pm i \epsilon}
\label{eq:CX}
\eeq
to any pre-determined precision.  

In applications the integral equation is approximated by projecting on
an approximation space ${\cal V}_k$ with a finest resolution $k$
dictated by the problem.  This projection can be computed efficiently
in the scaling basis (\ref{eq:CR}) using the one-point quadrature
(\ref{eq:CW}) and the explicit integrals (\ref{eq:CX}).  The resulting
matrix equation is transformed using the wavelet transform to an
equivalent system in the wavelet basis (\ref{eq:CS}).  In the
transformed basis the kernel of the integral equation decomposes into
the sum of a sparse matrix and a small matrix.  The kernel is
approximated by setting matrix elements of the kernel that are smaller
than a threshold value to zero.  This results in a sparse matrix
approximation.  The resulting linear system is solved using sparse
matrix iterative techniques, such as the complex biconjugate gradient
method \cite{numrec}\cite{golub} used in \cite{kessler04}.  This
solution is transformed back to the scaling function representation,
using the inverse wavelet transform, and the resulting solution is
inserted back in the integral equation to construct an interpolated
solution \cite{Sloan}.

The only wavelet information used in this application is the wavelet
transform and moments of the scaling function.  These can both be
expressed directly in terms of the scaling coefficients $h_l$ in Table
1.  This means that the basis functions never have to be computed.

The work in references \cite{kessler03} \cite{kessler04} shows that
all of these steps work as expected.  These references also discuss
technical issues that arise due to the treatment of endpoints when the
equations are transformed to a finite interval. 

\section{Dynamical Equations}

The general structure of the Faddeev-Lovelace
\cite{fritz65}\cite{wp91} equation in a relativistic quantum theory
with three particles of mass $m$ is
\[
X (p,q;p',q') = D  (p,q;p',q')+
\]
\beq
\int_0^{\infty} {K (p,q;p'',q'',z)   dp'' dq'' \over
z - e_1 (p'',q'') - e_2 (q'') } X (p'',q'';p',q')   
\label{eq:DA}
\eeq
where 
\beq
e_1 (p,q) = \sqrt{4 p^2 + 4m^{2} + q^{2}}
\qquad  
e_2 (q) = \sqrt{m^{2} + q^{2}}
\label{eq:DB}
\eeq
and $X=X_{mn}$, $K=K_{mn}$ and $D=D_{mn}$ are complex matrix valued
functions.  The quantity $K(p,q;p',q',z)$ is the smooth part of the
kernel.

In order to use wavelet methods it is advantageous to transform this
equation to a form where functions of the momentum, rather than the
energy, are additive in the denominator.  This transformation
facilitates the treatment of the moving singularity.  Note that $e_1>
e_2$ for all values of $p$ and $q$.  If $E>0$ then $E+ e_1(p,q) - e_2
(q) >0$.  It follows that the singular denominator
\beq
{1 \over E + i 0^+ - e_1 (p,q) - e_2 (q)}, 
\label{eq:DC}
\eeq
where $z=E+i 0^+$,  can be transformed to a more useful form 
by multiplying the 
numerator and denominator by the non-zero function 
$E+ e_1(p,q) - e_2 (q)$. This leads to the equivalent expression
\[
{1 \over E + i 0^+ - e_1 - e_2 }  =
\]
\[
{E+ e_1 - e_2  \over 
E^2 + e_2^2 - 2 E e_2 -e_1^2 + i 0^+ (E+ e_1 - e_2) } =
\]
\beq
{E+ e_1 - e_2  \over 
E^2-3m^2 - 2 E \sqrt{q^2 + m^2}  - 4 p^2  + i 0^+ } 
\label{eq:DD}
\eeq
which has the advantage that it separates the $p$ and $q$ dependence. 
In this expression there is only one singularity in the denominator.
%The coefficient of $i0^+$ is positive, but momentum dependent. 
%It can be shown that (\ref{eq:DD}) is equivalent to:
%\beq
%{E+ e_1 - e_2   \over 
%E^2-3m^2 - 2 E \sqrt{q^2 + m^2}  -4 p^2  + i 0^+  }. 
%\label{eq:DE}
%\eeq
The next step is to change variables
\beq
x= \eta 4 p^2
\qquad
y = \eta 2 E (\sqrt{q^2 + m^2}-m)  
\label{eq:DF}
\eeq
and define
\beq
z' = \eta [(E-m)^2 - 4m^2] .
\label{eq:DG}
\eeq
The parameter $\eta$ both sets a scale and can be used to fine tune
$z'$ so the real part is a dyadic rational of the form $n/2^{-k}$.
The method that we use to evaluate the singular integrals requires that 
$z$ is a dyadic rational.

The substitutions (\ref{eq:DF}) and (\ref{eq:DG}) 
		   lead to the equivalent equation
\[
\bar{X} (x,y;x',y') = \bar{D}  (x,y;x',y')+
\]
\beq
\int_0^{\infty}  {\bar{K} (x,y;x'',y'')dx'' dy'' \over
z' - x'' -y'' + i 0^+ } \bar{X} (x'',y'';x',y')   
\label{eq:DH}
\eeq
where
\beq
\bar{X} (x,y;x',y') = {X} (p(x),q(y);p(x'),q(y')) 
\label{eq:DI}
\eeq
\beq
\bar{D}  (x,y;x',y') = {D} (p(x),q(y);p(x'),q(y'))
\label{eq:DJ}
\eeq
\[
\bar{K} (x,y;x',y') = 
{K} (p(x),q(y);p(x'),q(y'),z) \times 
\]
\beq
\eta (E+ e_1(p(x'),q(y')) - e_2 (q(y'))) 
\vert {dp \over dx'}{dq \over dy'} \vert . 
\label{eq:DK}
\eeq

Approximate equations are derived using projection methods.  We seek a
solution $\bar{X}$, in the $x$, and $y$ variables, on the
approximation space ${\cal V}_k \times {\cal V}_k$.  Approximate equations
are obtained by projecting the smooth part of the kernel and the 
driving on this space. We use the approximations:
\[
\bar{X} (x,y;x',y') \approx
\]
\beq
\sum_{mn} \phi_{km} (x) \phi_{kn}(y) \bar{X}_{m,n}(x',y'),  
\label{eq:DL}
\eeq
\[
\bar{D}  (x,y;x',y')\approx 
\]
\beq
\sum_{mn} 
\phi_{km} (x) \phi_{kn}(y)
\bar{D}_{m,n}(x',y'),
\label{eq:DM}
\eeq
and
\[
\bar{K}  (x,y;x',y')\approx 
\]
\beq
\sum_{mnm'n'} \phi_{km} (x) \phi_{kn}(y) \bar{K}_{m,n;m',n'} 
\phi_{km'} (x') \phi_{kn'}(y')
\label{eq:DN}
\eeq
where 
\beq
\bar{D}_{m,n}(x',y'):= 2^k \bar{D}(x_m,x_n,x',y')
\eeq
\beq
\bar{K}_{m,n;m',n'}:= 2^{2k} \bar{K}(x_m,x_n;x_{m'},x_{n'})
\eeq
are evaluated at the one point quadrature points associated with 
$\phi_{km}(x)$:  
\beq
x_m = 2^{k}(<x>+m) \qquad <x>= {1 \over \sqrt{2}}\sum_{l=1}^{2K-1} l h_l.
\label{eq:DNA}
\eeq
It is useful to approximate
the product of the approximate expressions, (\ref{eq:DL}) and
(\ref{eq:DN}) for $\bar{K} (x,y;x',y')$ and $\bar{X} (x',y')$ by
expanding the $x'$ and $y'$ dependence in the basis on ${\cal V}_k
\times {\cal V}_k $.  The justification for this approximation is that
if both of the approximations are well represented by low-degree
polynomials on the scale $k$ in $x'$ and $y'$, then the product of
these functions should be well represented by low-degree polynomials
on the scale $k$ in $x'$ and $y'$.

To test this approximation we approximate $x^4$ by re-expanding the
product of expansions of $x^2$ using Daubechies' $K=3$ wavelets. 
We write
\beq
x^2 = \sum_n x_n^2 \phi_{kn}(x)
\eeq
which gives
\beq
x^4= \sum_{mn} x_m^2 x_n^2 \phi_{km}(x) \phi_{kn}(x) .
\eeq
This is exact for the Daubechies' $K=3$ wavelets.  
We approximate this by projecting on the approximation 
space ${\cal V}_k$.  The expansion coefficients are
\beq
c_l = \int x^4 \phi_{kl}(x) dx= 
\sum_{mn}  x_m^2 x_n^2 I^k_{mnl}
\eeq
where 
\beq
I^{k}_{l m n} := \int 
\phi_{kl} (x) \phi_{km}(x) \phi_{kn}(x) dx .
\label{eq:DQ}
\eeq
The approximation defined by this projection can be written as
\beq
x^4 \approx  \sum_{mnl}  x_m^2 x_n^2 I^k_{mnl}\phi_{kl}(x)
\label{eq:DPA}
\eeq

The expansion coefficients $x_n^2$  are computed using the 1 point
quadrature, which is exact for the expansion of $x^2$.  The
$\phi_{kl}(x)$ are evaluated at dyadic rationals so there is no error
in computing the scaling basis functions.  The only
source of error is the approximation (\ref{eq:DPA}).  Table 2 compares
the right and left sides of equation (\ref{eq:DPA}) for resolution 
$k=-5$ 

\begin{center}
\begin{table} % [hbt]
{\bf Table 2: Test of double expansion } \\[1.0ex]
\begin{tabular}{ c c c }
\hline
\hline
$x$ & $x^4$ & $\sum x_m^2 x_n^2 I_{mnk} \phi_k$  \\  
\hline					     		      
 $-1.000000e+01$&$1.000000e+04$&$ 1.000000e+04$ \\
%-9.875000e+00& 9.509297e+03& 9.509298e+03	\\
%-9.750000e+00& 9.036879e+03& 9.036879e+03	\\
%-9.625000e+00& 8.582285e+03& 8.582286e+03	\\
%-9.500000e+00& 8.145062e+03& 8.145063e+03	\\
%-9.375000e+00& 7.724762e+03& 7.724762e+03	\\
%-9.250000e+00& 7.320941e+03& 7.320942e+03	\\
%-9.125000e+00& 6.933164e+03& 6.933165e+03	\\
 $-9.000000e+00$&$6.561000e+03$&$6.561000e+03$	\\
%-8.875000e+00& 6.204024e+03& 6.204024e+03	\\
%-8.750000e+00& 5.861816e+03& 5.861817e+03	\\
%-8.625000e+00& 5.533965e+03& 5.533966e+03	\\
%-8.500000e+00& 5.220062e+03& 5.220063e+03	\\
%-8.375000e+00& 4.919707e+03& 4.919708e+03	\\
%-8.250000e+00& 4.632504e+03& 4.632504e+03	\\
%-8.125000e+00& 4.358063e+03& 4.358063e+03	\\
 $-8.000000e+00$&$4.096000e+03$&$4.096000e+03$	\\
%-7.875000e+00& 3.845938e+03& 3.845938e+03	\\
%-7.750000e+00& 3.607504e+03& 3.607504e+03	\\
%-7.625000e+00& 3.380332e+03& 3.380333e+03	\\
%-7.500000e+00& 3.164062e+03& 3.164063e+03	\\
%-7.375000e+00& 2.958340e+03& 2.958340e+03	\\
%-7.250000e+00& 2.762816e+03& 2.762817e+03	\\
%-7.125000e+00& 2.577149e+03& 2.577149e+03	\\
$-7.000000e+00$&$2.401000e+03$&$2.401000e+03$	\\
%-6.875000e+00& 2.234039e+03& 2.234040e+03	\\
%-6.750000e+00& 2.075941e+03& 2.075942e+03	\\
%-6.625000e+00& 1.926387e+03& 1.926387e+03	\\
%-6.500000e+00& 1.785062e+03& 1.785063e+03	\\
%-6.375000e+00& 1.651660e+03& 1.651661e+03	\\
%-6.250000e+00& 1.525879e+03& 1.525879e+03	\\
%-6.125000e+00& 1.407422e+03& 1.407422e+03	\\
 $-6.000000e+00$&$1.296000e+03$&$1.296000e+03$	\\
%-5.875000e+00& 1.191328e+03& 1.191329e+03	\\
%-5.750000e+00& 1.093129e+03& 1.093129e+03	\\
%-5.625000e+00& 1.001129e+03& 1.001129e+03	\\
%-5.500000e+00& 9.150625e+02& 9.150628e+02	\\
%-5.375000e+00& 8.346682e+02& 8.346685e+02	\\
%-5.250000e+00& 7.596914e+02& 7.596917e+02	\\
%-5.125000e+00& 6.898831e+02& 6.898833e+02	\\
 $-5.000000e+00$&$6.250000e+02$&$6.250002e+02$	\\
%-4.875000e+00& 5.648049e+02& 5.648052e+02	\\
%-4.750000e+00& 5.090664e+02& 5.090666e+02	\\
%-4.625000e+00& 4.575588e+02& 4.575591e+02	\\
%-4.500000e+00& 4.100625e+02& 4.100627e+02	\\
%-4.375000e+00& 3.663635e+02& 3.663637e+02	\\
%-4.250000e+00& 3.262539e+02& 3.262541e+02	\\
%-4.125000e+00& 2.895315e+02& 2.895317e+02	\\
 $-4.000000e+00$&$2.560000e+02$&$2.560002e+02$	\\
%-3.875000e+00& 2.254690e+02& 2.254692e+02	\\
%-3.750000e+00& 1.977539e+02& 1.977541e+02	\\
%-3.625000e+00& 1.726760e+02& 1.726762e+02	\\
%-3.500000e+00& 1.500625e+02& 1.500627e+02	\\
%-3.375000e+00& 1.297463e+02& 1.297465e+02	\\
%-3.250000e+00& 1.115664e+02& 1.115666e+02	\\
%-3.125000e+00& 9.536743e+01& 9.536758e+01	\\
 $-3.000000e+00$&$8.100000e+01$&$8.100015e+01$	\\
%-2.875000e+00& 6.832056e+01& 6.832070e+01	\\
%-2.750000e+00& 5.719141e+01& 5.719154e+01	\\
%-2.625000e+00& 4.748071e+01& 4.748084e+01	\\
%-2.500000e+00& 3.906250e+01& 3.906262e+01	\\
%-2.375000e+00& 3.181665e+01& 3.181677e+01	\\
%-2.250000e+00& 2.562891e+01& 2.562902e+01	\\
%-2.125000e+00& 2.039087e+01& 2.039097e+01	\\
 $-2.000000e+00$&$1.600000e+01$&$1.600010e+01$	\\  
%-1.875000e+00& 1.235962e+01& 1.235971e+01	\\
%-1.750000e+00& 9.378906e+00& 9.378993e+00	\\
%-1.625000e+00& 6.972900e+00& 6.972981e+00	\\
 $-1.500000e+00$&$5.062500e+00$&$5.062574e+00$	\\
%-1.375000e+00& 3.574463e+00& 3.574531e+00	\\
%-1.250000e+00& 2.441406e+00& 2.441468e+00	\\
%-1.125000e+00& 1.601807e+00& 1.601863e+00	\\
 $-1.000000e+00$&$1.000000e+00$&$1.000050e+00$	\\
%-8.750000e-01& 5.861816e-01& 5.862257e-01	\\
%-7.500000e-01& 3.164062e-01& 3.164443e-01	\\
%-6.250000e-01& 1.525879e-01& 1.526199e-01	\\
 $-5.000000e-01$&$6.250000e-02$&$6.252593e-02$	\\
 $-3.750000e-01$&$1.977539e-02$&$1.979528e-02$	\\
 $-2.500000e-01$&$3.906250e-03$&$3.920094e-03$	\\
 $-1.250000e-01$&$2.441406e-04$&$2.519414e-04$	\\
 $\phantom{-}0.000000e+00$&$0.000000e+00$&$1.757732e-06$	\\
 $\phantom{-}1.250000e-01$&$2.441406e-04$&$2.398553e-04$	\\
 $\phantom{-}2.500000e-01$&$3.906250e-03$&$3.895922e-03$	\\
 $\phantom{-}3.750000e-01$&$1.977539e-02$&$1.975902e-02$	\\
 $\phantom{-}5.000000e-01$&$6.250000e-02$&$6.247759e-02$	\\
%6.250000e-01&  1.525879e-01& 1.525594e-01	\\
%7.500000e-01&  3.164062e-01& 3.163717e-01	\\
%8.750000e-01&  5.861816e-01& 5.861411e-01	\\
 $\phantom{-}1.000000e+00$& $1.000000e+00$&$9.999534e-01$	\\
%1.125000e+00&  1.601807e+00& 1.601754e+00	\\
%1.250000e+00&  2.441406e+00& 2.441348e+00	\\
%1.375000e+00&  3.574463e+00& 3.574398e+00	\\
%1.500000e+00&  5.062500e+00& 5.062429e+00	\\
%1.625000e+00&  6.972900e+00& 6.972824e+00	\\
%1.750000e+00&  9.378906e+00& 9.378823e+00	\\
%1.875000e+00&  1.235962e+01& 1.235953e+01	\\
 $\phantom{-}2.000000e+00$&$1.600000e+01$&$1.599991e+01$	\\
%2.125000e+00&  2.039087e+01& 2.039077e+01	\\
%2.250000e+00&  2.562891e+01& 2.562880e+01	\\
%2.375000e+00&  3.181665e+01& 3.181654e+01	\\
%2.500000e+00&  3.906250e+01& 3.906238e+01	\\
%2.625000e+00&  4.748071e+01& 4.748059e+01	\\
%2.750000e+00&  5.719141e+01& 5.719128e+01	\\
%2.875000e+00&  6.832056e+01& 6.832042e+01	\\
 $\phantom{-}3.000000e+00$&$8.100000e+01$&$8.099986e+01$	\\
%3.125000e+00&  9.536743e+01& 9.536728e+01	\\
%3.250000e+00&  1.115664e+02& 1.115663e+02	\\
%3.375000e+00&  1.297463e+02& 1.297462e+02	\\
%3.500000e+00&  1.500625e+02& 1.500623e+02	\\
%3.625000e+00&  1.726760e+02& 1.726759e+02	\\
%3.750000e+00&  1.977539e+02& 1.977537e+02	\\
%3.875000e+00&  2.254690e+02& 2.254688e+02	\\
 $\phantom{-}4.000000e+00$&$2.560000e+02$&$2.559998e+02$	\\
%4.125000e+00&  2.895315e+02& 2.895313e+02	\\
%4.250000e+00&  3.262539e+02& 3.262537e+02	\\
%4.375000e+00&  3.663635e+02& 3.663633e+02	\\
%4.500000e+00&  4.100625e+02& 4.100623e+02	\\
%4.625000e+00&  4.575588e+02& 4.575586e+02	\\
%4.750000e+00&  5.090664e+02& 5.090662e+02	\\
%4.875000e+00&  5.648049e+02& 5.648047e+02	\\
 $\phantom{-}5.000000e+00$&$6.250000e+02$&$6.249998e+02$	\\
%5.125000e+00&  6.898831e+02& 6.898828e+02	\\
%5.250000e+00&  7.596914e+02& 7.596912e+02	\\
%5.375000e+00&  8.346682e+02& 8.346680e+02	\\
%5.500000e+00&  9.150625e+02& 9.150622e+02	\\
%5.625000e+00&  1.001129e+03& 1.001129e+03	\\
%5.750000e+00&  1.093129e+03& 1.093129e+03	\\
%5.875000e+00&  1.191328e+03& 1.191328e+03	\\
 $\phantom{-}6.000000e+00$&$1.296000e+03$&$1.296000e+03$	\\  
%6.125000e+00&  1.407422e+03& 1.407422e+03	\\
%6.250000e+00&  1.525879e+03& 1.525879e+03	\\
%6.375000e+00&  1.651660e+03& 1.651660e+03	\\
%6.500000e+00&  1.785062e+03& 1.785062e+03	\\
%6.625000e+00&  1.926387e+03& 1.926387e+03	\\
%6.750000e+00&  2.075941e+03& 2.075941e+03	\\
%6.875000e+00&  2.234039e+03& 2.234039e+03	\\
 $\phantom{-}7.000000e+00$&$2.401000e+03$&$2.401000e+03$\\
%7.125000e+00&  2.577149e+03& 2.577148e+03	\\
%7.250000e+00&  2.762816e+03& 2.762816e+03	\\
%7.375000e+00&  2.958340e+03& 2.958340e+03	\\
%7.500000e+00&  3.164062e+03& 3.164062e+03	\\
%7.625000e+00&  3.380332e+03& 3.380332e+03	\\
%7.750000e+00&  3.607504e+03& 3.607504e+03	\\
%7.875000e+00&  3.845938e+03& 3.845937e+03	\\
 $\phantom{-}8.000000e+00$&$4.096000e+03$&$4.096000e+03$	\\
%8.125000e+00&  4.358063e+03& 4.358062e+03	\\
%8.250000e+00&  4.632504e+03& 4.632504e+03	\\
%8.375000e+00&  4.919707e+03& 4.919707e+03	\\
%8.500000e+00&  5.220062e+03& 5.220062e+03	\\
%8.625000e+00&  5.533965e+03& 5.533965e+03	\\
%8.750000e+00&  5.861816e+03& 5.861816e+03	\\
%8.875000e+00&  6.204024e+03& 6.204023e+03	\\
 $\phantom{-}9.000000e+00$&$6.561000e+03$&$6.561000e+03$	\\
%9.125000e+00&  6.933164e+03& 6.933164e+03	\\
%9.250000e+00&  7.320941e+03& 7.320941e+03	\\
%9.375000e+00&  7.724762e+03& 7.724762e+03	\\
%9.500000e+00&  8.145062e+03& 8.145062e+03	\\
%9.625000e+00&  8.582285e+03& 8.582285e+03	\\
%9.750000e+00&  9.036879e+03& 9.036878e+03	\\
%9.875000e+00&  9.509297e+03& 9.509297e+03	\\
 $\phantom{-}1.000000e+01$&$1.000000e+04$&$1.000000e+04$	\\
\hline
\hline
\end{tabular}
% \label{tabspec}
\end{table}
\end{center}
The expansion is essentially exact, except near the critical point,
$x^4=0$, where it is still accurate.  The accuracy near the critical
point can be improved using a higher resolution, however a degenerate 
critical point is not generic.

This additional approximation gives 
\[
\bar{K}
(x,y;x',y') \bar{X} (x',y';x'',y'')
\approx
\]
\[
\sum \phi_{km} (x) \phi_{kn}(y) \bar{K}_{m,n;m',n'} 
I^k_{n' n'' n'''}
I^k_{m' m'' m'''} \times
\]
\beq
\bar{X}_{m'',n''}(x'',y'')  
\phi_{km'''} (x') \phi_{k n'''}(y').
\label{eq:DP}
\eeq
Even though this introduces two additional sums, most of the terms are
zero because $I^k_{m, m',m''}=0$ unless $\vert m-m' \vert$,
$\vert m'-m''\vert$ and $\vert m''-m \vert$ are all less than $2K-1$.  
In section
IV. we show that the integrals $I^k_{m m' m''}$ can all be computed
analytically using the scaling equation.

With these approximations the dynamical equations reduce to the algebraic 
system  
\[
\bar{X}_{m,n}(x',y') = \bar{D}_{m,n}(x',y')+
\]
\beq
\sum \bar{K}_{m,n;m',n'} I^k_{m''' m' m''} 
I^k_{n''' n' n''} J^k_{m''',n'''}(z)
\bar{X}_{m'',n''}(x',y')   
\label{eq:DR}
\eeq
where
\beq
J^k_{m,n}(z) :=
\int_0^{\infty} dx dy {\phi_{km} (x) 
\phi_{kn} (y) \over
z - x -y + i 0^+ }.
\label{eq:DS}
\eeq

These equations separate the smooth part of the physics input in
$\bar{D}$ and $\bar{K}$ from the singular part of this equation,
contained in the integrals $J^k_{mn}(z)$. While the construction of
the smooth kernel in the relativistic case is considerably more
complicated than in the non-relativistic case \cite{wp91}\cite{fritz65}, 
given the
driving term and smooth kernel the projections $\bar{K}_{m,n;m',n'}$
and $\bar{D}_{ m,n}(x',y')$ can be calculated by evaluating the exact
driving term and kernel at the one-point quadrature point for each
$\phi_{km} (x)$.  This reduces a Galerkin projection to a simple
function evaluation.

In the next section we discuss the evaluation of the integrals 
\beq
I^{k}_{l,m,n}
\qquad
\mbox{and} 
\qquad
J^k_{m,n}(z)
\label{eq:DT}
\eeq
that appear in (\ref{eq:DR}).  These integrals can be evaluated and
stored before calculation.  They are the wavelet input to the
calculation.  They replace all of the integrations in the integral
equations and they are {\it independent} of the choice of dynamical
model.  The physics input is contained in the matrices $\bar{K}_{m,n;m',n'}$
and $\bar{D}_{mn}(x,y)$.  Equation (\ref{eq:DR}) gives a clean and
stable separation of the physics and the treatment of the moving
singularity, which is contained in the integrals (\ref{eq:DT}).

The equations (\ref{eq:DR}) are an infinite set of equations.  They
can be reduced to a finite set by including high-momentum cutoffs or
transforming to a finite interval.  The treatment of endpoints in the
evaluations of $\bar{K}_{m,n;m',n'}$ and $\bar{D}_{m,n}(x',y')$ is
identical to the treatment used in \cite{kessler03} and
\cite{kessler04}, where partial moments of the scaling function are
used to construct simple quadratures that exactly integrate the
product of the scaling function and degree $K-1$ polynomials over a
subinterval of the support of the scaling function
\cite{Shann1}\cite{Shann2}\cite{Sweldens} .  The treatment of
endpoints in the evaluation of $I^{m}_{n k}$ and $J_{m,n}(z)$ is
discussed in this paper.

Even with the reduction to a finite set of equations, the system of
equations is large.  It can be reduced by performing a wavelet
transform on the scaling function basis.  This can be done following
the method used in \cite{kessler04}\cite{numrec}, which maps the
interval to a circle to treat endpoints.  This does not change
the final result because the resulting transformation is still
a finite orthogonal transformation.

The next step is to discard the small matrix elements in the
transformed kernel and to solve the resulting equation.

As discovered in \cite{kessler04}, the treatment of the endpoints
leads to an ill-conditioned matrix.  This is because the right tail of
the scaling function is small (see Fig. 1).  Some of the overlap
integrals with support containing the left endpoint replace the
orthogonality integrals by integrals of products of scaling
functions over an interval where the product is small.  This can be
fixed using the conditioning method that was used in \cite{kessler04}.
The resulting conditioned equations are stable and can be accurately
solved using sparse matrix techniques.

The resulting solution can be transformed back to the scaling function 
basis.   An interpolated solution is then constructed from
the solution, $\bar{X}_{m,n}(x,y)$, of the algebraic equations
using the Sloan interpolation method \cite{Sloan}
\[
\bar{X} (x,y;x',y') = \bar{D}  (x,y;x',y')+
\]
\beq
\sum \bar{K}_{ m,n} (x,y) I^k_{m m'm''} 
I^k_{n n'n''} J^k_{m'',n''}(z)
\bar{X}_{m',n'}(x',y').   
\label{eq:DU}
\eeq
If this interpolation is used the basis functions never have to be evaluated.

Equations (\ref{eq:DR}) and (\ref{eq:DU}) along with the methods for 
computing the integrals (\ref{eq:DT}) are the main results of this 
paper.

\section{ Evaluation of Integrals}

In this section we discuss the evaluation of the integrals 
$I^k_{l,n,m}$ and  $J^k_{mn} (z)$ that appear in equation
(\ref{eq:DR}).  These integrals are defined in equations 
(\ref{eq:DQ}) and (\ref{eq:DS}).

To evaluate these integrals we first express the scale ``$k$''
integrals in terms of the scale ``$0$'' integrals, then we
evaluate the scale ``$0$'' integrals.  Using the
definition (\ref{eq:CJ}) in equations (\ref{eq:DR}) and (\ref{eq:DS})
we obtain
\beq
I^k_{l,n,m} 
= 2^{-k/2} I^0_{l,n,m}
\label{eq:H4}
\eeq
and 
\beq
J^k_{mn} (z) =
J^0_{mn} (2^{-k}z).
\label{eq:H5}
\eeq 
For $k$ a negative integer, we can choose $2^{-k}z$ as an integer
which is equivalent to choosing $z$ to be a dyadic rational.  This can
be done for any on shell energy by adjusting the parameter $\eta$ in
(\ref{eq:DF}).  As a result, it is enough to evaluate $J^0_{m,n} (l)$
and $I^0_{l,m,n}$
for $l,m,n$ integers.  In what follows we define
\beq
I_{l,m,n} := I^0_{l,m,n}
\label{eq:H71}
\eeq
and
\beq
J_{m,n} (k):= J^0_{m,n}(k).
\label{eq:H72}
\eeq
Both $I_{l,n,m}$ and $J_{m,n} (k)$ involve integrals over the half
infinite interval.  In order to evaluate these integrals we first
evaluate the corresponding integrals over the infinite interval:
\beq
\bar I_{l,n,m} = \int_{-\infty}^{\infty} \phi (x-l) \phi (x-n) \phi (x-m) dx 
\label{eq:H8}
\eeq
and
\beq
\bar J_{m,n} (k) = 
\int_{-\infty}^{\infty} dx \int_{-\infty}^{\infty} 
dy {\phi (x-m) \phi (y-n) \over 
k- x- y + i0^+}. 
\label{eq:H9}
\eeq
These integrals are easier to compute because of the simplified boundary 
conditions.  

The support of the scaling functions implies that 
if any of $l$, $m$ or $n$ are non-negative then
\beq
I_{l,m,n} = \bar I_{l,m,n},
\label{eq:H10}
\eeq 
and if any of $l$,$m$ or $n$ are less than $-2K+2$ then
\beq
I_{l,m,n} = 0.
\label{eq:H11}
\eeq 
The non-trivial values of $I_{l,m,n}$ correspond to the case that 
the indices $l$, $m$, $n$ satisfy
\beq
l,m,n \in [-2K+2,-1] .
\label{eq:H12}
\eeq
To compute the integrals $\bar{I}_{l,m,n}$ defined in (\ref{eq:H8}) note 
that the definition implies
\beq
\bar I_{l,m,n} = \bar I_{0,m-l,n-l} .
\label{eq:H14}
\eeq
which allows us to express $\bar I_{l,m,n}$ in terms of  
$\bar I_{n,m}$ defined by 
\beq
\bar I_{m,n} = I_{0,m,n}= 
\int_{-\infty}^{\infty} dx \phi (x) \phi (x-m) \phi (x-n) .
\label{eq:H15}
\eeq
Since the support of $\phi (x)$ is contained in the interval
$[0,2K-1]$, there are only a finite number of non-zero values of 
$\bar I_{m,n}$.  These have $
m,n\in [-2K+2, 2K-2]$.  For $K=3$ there are 81 non-zero $\bar
I_{m,n}$ with $m,n\in [-4,4]$.   

We can derive linear equations relating these integrals 
using the scaling equation in the form
\beq
\phi (x) = \sqrt{2} \sum_{l=0}^{2K-1} h_l \phi (2x-l). 
\label{eq:H6}
\eeq

When we use (\ref{eq:H6}) in equation (\ref{eq:H15}), 
the resulting scaling equations for the integrals $\bar{I}_{mn}$ are:
\beq
\bar I_{m,n} = \sqrt{2} 
\sum_{l_m,l_n,l_k=0}^{2K-1} h_{l_k} h_{l_m} h_{l_n} 
\bar{I}_{2m+l_m-l_k,2n+l_n-l_k}.
\label{eq:H16}
\eeq
These are homogeneous equations relating the non-zero values of 
$\bar{I}_{m,n}$.  An additional  
inhomogeneous equation is needed to solve for the non-zero
values of $\bar{I}_{m,n}$.  The needed equation
follows from the normalization 
condition (\ref{eq:CB}) and the identity
\beq
\sum_n \phi (x-n) =1
\label{eq:H16a}
\eeq
which when used in (\ref{eq:H15}) gives  
the inhomogeneous equation
\beq
\sum_{m=-2K+2}^{2K-2}  \bar I_{m,n} = \delta_{n0}.
\label{eq:H17}
\eeq
Equations (\ref{eq:H16}) and (\ref{eq:H17}) are a finite system 
of $(4K-3)\times (4K-3)$ linear equations that can be 
solved for the non-zero values of $\bar{I}_{mn}$.
The results of these calculations $\bar{I}_{mn}$ for $K=3$ 
are given in Table 3. 

These solutions give $I_{lmn}$ when $l,m$ or $n$ are non-negative 
from equations (\ref{eq:H10}) and (\ref{eq:H14}).  To calculate remaining 
non-zero values of $I_{lmn}$ first observe that using (\ref{eq:H6}) in 
(\ref{eq:DQ}) gives    
scaling equations for $I_{k,m,n}$:
\beq
I_{k,m,n} = \sqrt{2} \sum h_{l_k} h_{l_m} h_{l_n} I_{2k+l_k,2m+l_m,2n+l_n}.
\label{eq:H7}
\eeq
These equations are not homogeneous equations because when any of the
indices on the right hand side of the equation are non-negative,
$I_{k,m,n} = \bar{I}_{k,m,n}= \bar{I}_{m-k,n-k}$, which is known
input.  This linear system can be solved for the non trivial values of
$I_{k,m,n}$ associated with the values of $k,n,m \in [-2K+2,-1]$.
For $K=3$ there are 64 values of $k,m,n \in [-4,-1]$.  The
results of this calculation for the $K=3$ case are given in Table 4.
                                                     
\begin{table} % [hbt]
{\bf Table 3 - $\bar{I}_{mn}$} \\[1.0ex]
\begin{tabular}{ c c c c c c }
%\firsthline
\hline
\hline
m & n & $\bar{I}_{mn}$ & m & n & $\bar{I}_{mn}$ \\ 
\hline					    		      
-4& -4&$\phantom{-}1.160637e-07$  & \phantom{-}1 &\phantom{-}0 &$\phantom{-}1.469238e-01$  \\   
-3& -4&$\phantom{-}9.788805e-07$  & \phantom{-}2 &\phantom{-}0 &$\phantom{-}7.027929e-03$  \\ 
-2& -4&$-2.811543e-06$ & \phantom{-}3 &\phantom{-}0 &$\phantom{-}2.025919e-04$  \\ 
-1& -4&$\phantom{-}6.184412e-06$  & \phantom{-}4 &\phantom{-}0 &$\phantom{-}1.160637e-07$  \\ 
\phantom{-}0 & -4&$-4.467813e-06$ &  -4&\phantom{-}1 &$\phantom{-}0.000000e+00$  \\ 
\phantom{-}1 & -4&$\phantom{-}0.000000e+00$  &  -3&\phantom{-}1 &$\phantom{-}6.184412e-06$  \\ 
\phantom{-}2 & -4&$\phantom{-}0.000000e+00$  &  -2&\phantom{-}1 &$\phantom{-}1.159627e-03$  \\ 
\phantom{-}3 & -4&$\phantom{-}0.000000e+00$  &  -1&\phantom{-}1 &$-3.047012e-02$ \\ 
\phantom{-}4 & -4&$\phantom{-}0.000000e+00$  &  \phantom{-}0 &\phantom{-}1 &$\phantom{-}1.469238e-01$  \\ 
-4& -3&$\phantom{-}9.788805e-07$  & \phantom{-}1&\phantom{-}1 &$-8.660587e-02$ \\ 
-3& -3&$\phantom{-}2.025919e-04$  & \phantom{-}2&\phantom{-}1 &$-3.047012e-02$ \\ 
-2& -3&$-5.444572e-04$ & \phantom{-}3&\phantom{-}1 &$-5.444572e-04$ \\ 
-1& -3&$\phantom{-}1.159627e-03$  & \phantom{-}4 &\phantom{-}1 &$\phantom{-}9.788805e-07$  \\ 
\phantom{-}0 & -3&$-8.249248e-04$ &  -4& \phantom{-}2 &$\phantom{-}0.000000e+00$  \\ 
\phantom{-}1 & -3&$\phantom{-}6.184412e-06$  &  -3& \phantom{-}2 &$\phantom{-}0.000000e+00$  \\ 
\phantom{-}2 & -3&$\phantom{-}0.000000e+00$  &  -2& \phantom{-}2 &$-2.811543e-06$ \\ 
\phantom{-}3 & -3&$\phantom{-}0.000000e+00$  &  -1& \phantom{-}2 &$-5.444572e-04$ \\ 
\phantom{-}4 & -3&$\phantom{-}0.000000e+00$  &  \phantom{-}0 & \phantom{-}2 &$\phantom{-}7.027929e-03$  \\ 
-4& -2&$-2.811543e-06$ &  \phantom{-}1 & \phantom{-}2 &$-3.047012e-02$ \\ 
-3& -2&$-5.444572e-04$ &  \phantom{-}2 & \phantom{-}2 &$\phantom{-}2.283264e-02$  \\ 
-2& -2&$\phantom{-}7.027929e-03$  &  \phantom{-}3 & \phantom{-}2 &$\phantom{-}1.159627e-03$  \\ 
-1& -2&$-3.047012e-02$ &  \phantom{-}4 & \phantom{-}2 &$-2.811543e-06$ \\ 
\phantom{-}0 & -2&$\phantom{-}2.283264e-02$  &  -4& \phantom{-}3 &$\phantom{-}0.000000e+00$  \\ 
\phantom{-}1 & -2&$\phantom{-}1.159627e-03$  &  -3& \phantom{-}3 &$\phantom{-}0.000000e+00$  \\ 
\phantom{-}2 & -2&$-2.811543e-06$ &  -2& \phantom{-}3 &$\phantom{-}0.000000e+00$  \\ 
\phantom{-}3 & -2&$\phantom{-}0.000000e+00$  &  -1& \phantom{-}3 &$\phantom{-}9.788805e-07$  \\ 
\phantom{-}4 & -2&$\phantom{-}0.000000e+00$  & \phantom{-}0 & \phantom{-}3 &$\phantom{-}2.025919e-04$  \\ 
-4& -1&$\phantom{-}6.184412e-06$  & \phantom{-}1 &\phantom{-}3 &$-5.444572e-04$ \\ 
-3& -1&$\phantom{-}1.159627e-03$  & \phantom{-}2 &\phantom{-}3 &$\phantom{-}1.159627e-03$  \\ 
-2& -1&$-3.047012e-02$ & \phantom{-}3 &\phantom{-}3 &$-8.249248e-04$ \\ 
-1& -1&$\phantom{-}1.469238e-01$  & \phantom{-}4 &\phantom{-}3 &$\phantom{-}6.184412e-06$  \\ 
\phantom{-}0 & -1&$-8.660587e-02$ &  -4& \phantom{-}4 &$\phantom{-}0.000000e+00$  \\ 
\phantom{-}1 & -1&$-3.047012e-02$ &  -3& \phantom{-}4 &$\phantom{-}0.000000e+00$  \\ 
\phantom{-}2 & -1&$-5.444572e-04$ &  -2& \phantom{-}4 &$\phantom{-}0.000000e+00$  \\ 
\phantom{-}3 & -1&$\phantom{-}9.788805e-07$  &  -1& \phantom{-}4 &$\phantom{-}0.000000e+00$  \\ 
\phantom{-}4 & -1&$\phantom{-}0.000000e+00$  & \phantom{-}0& \phantom{-}4 &$\phantom{-}1.160637e-07$  \\ 
-4& \phantom{-}0 &$-4.467813e-06$ & \phantom{-}1& \phantom{-}4 &$\phantom{-}9.788805e-07$  \\ 
-3& \phantom{-}0 &$-8.249248e-04$ & \phantom{-}2& \phantom{-}4 &$-2.811543e-06$ \\ 
-2& \phantom{-}0 &$\phantom{-}2.283264e-02$  & \phantom{-}3& \phantom{-}4 &$\phantom{-}6.184412e-06$  \\ 
-1& \phantom{-}0 &$-8.660587e-02$ & \phantom{-}4& \phantom{-}4 &$-4.467813e-06$ \\ 
\phantom{-}0 &\phantom{-} 0 &$\phantom{-}9.104482e-01$  &    &   &              \\          
%\lasthline
\hline
\hline		       
\end{tabular}					 
% \label{tabspec}
\end{table}
		        
\begin{table} % [hbt]
{\bf Table 4 - $\bar{I}_{mnl}$} \\[1.0ex]
\begin{tabular}{ c c c c c c c c }
%\firsthline
\hline
\hline
m & n & l &  $\bar{I}_{mnl}$ & m & n & l & $\bar{I}_{mnl}$ \\ 
\hline					    		      
-4&-4&-4&$\phantom{-}4.152357e-09$  &  -2&-4&-4&$-5.085054e-07$ \\ 
-4&-4&-3&$\phantom{-}1.155617e-07$&  -2&-4&-3&$-1.218375e-05$ \\ 
-4&-4&-2&$-5.085054e-07$ &  -2&-4&-2&$\phantom{-}5.118615e-05$  \\ 
-4&-4&-1&$\phantom{-}1.750639e-06$  &  -2&-4&-1&$-1.700711e-04$ \\ 
-4&-3&-4&$\phantom{-}1.155617e-07$ &  -2&-3&-4&$-1.218375e-05$ \\ 
-4&-3&-3&$\phantom{-}2.879737e-06$ &  -2&-3&-3&$-4.066737e-04$ \\ 
-4&-3&-2&$-1.218375e-05$ &  -2&-3&-2&$\phantom{-}1.869754e-03$  \\ 
-4&-3&-1&$\phantom{-}4.070309e-05$ &  -2&-3&-1&$-6.559712e-03$ \\ 
-4&-2&-4&$-5.085054e-07$&  -2&-2&-4&$\phantom{-}5.118615e-05$  \\ 
-4&-2&-3&$-1.218375e-05$ &  -2&-2&-3&$\phantom{-}1.869754e-03$  \\ 
-4&-2&-2&$\phantom{-}5.118615e-05$  &  -2&-2&-2&$-8.932389e-03$ \\ 
-4&-2&-1&$-1.700711e-04$ &  -2&-2&-1&$\phantom{-}3.218428e-02$  \\ 
-4&-1&-4&$\phantom{-}1.750639e-06$  &  -2&-1&-4&$-1.700711e-04$ \\ 
-4&-1&-3&$\phantom{-}4.070309e-05$&  -2&-1&-3&$-6.559712e-03$ \\ 
-4&-1&-2&$-1.700711e-04$ &  -2&-1&-2&$\phantom{-}3.218428e-02$  \\ 
-4&-1&-1&$\phantom{-}5.627612e-04$ &  -2&-1&-1&$-1.177691e-01$ \\ 
-3&-4&-4&$\phantom{-}1.155617e-07$  &  -1&-4&-4&$\phantom{-}1.750639e-06$  \\ 
-3&-4&-3&$\phantom{-}2.879737e-06$  &  -1&-4&-3&$\phantom{-}4.070309e-05$  \\ 
-3&-4&-2&$-1.218375e-05$ &  -1&-4&-2&$-1.700711e-04$ \\ 
-3&-4&-1&$\phantom{-}4.070309e-05$ &  -1&-4&-1&$\phantom{-}5.627612e-04$ \\ 
-3&-3&-4&$\phantom{-}2.879737e-06$  &  -1&-3&-4&$\phantom{-}4.070309e-05$ \\ 
-3&-3&-3&$\phantom{-}8.614462e-05$  &  -1&-3&-3&$\phantom{-}1.454880e-03$ \\ 
-3&-3&-2&$-4.066737e-04$ &  -1&-3&-2&$-6.559712e-03$ \\ 
-3&-3&-1&$\phantom{-}1.454880e-03$  &  -1&-3&-1&$\phantom{-}2.270045e-02$ \\ 
-3&-2&-4&$-1.218375e-05$ &  -1&-2&-4&$-1.700711e-04$\\ 
-3&-2&-3&$-4.066737e-04$ &  -1&-2&-3&$-6.559712e-03$ \\ 
-3&-2&-2&$\phantom{-}1.869754e-03$  &  -1&-2&-2&$\phantom{-}3.218428e-02$ \\ 
-3&-2&-1&$-6.559712e-03$ &  -1&-2&-1&$-1.177691e-01$ \\ 
-3&-1&-4&$\phantom{-}4.070309e-05$  &  -1&-1&-4&$\phantom{-}5.627612e-04$ \\ 
-3&-1&-3&$\phantom{-}1.454880e-03$  &  -1&-1&-3&$\phantom{-}2.270045e-02$  \\ 
-3&-1&-2&$-6.559712e-03$ &  -1&-1&-2&$-1.177691e-01$ \\ 
-3&-1&-1&$\phantom{-}2.270045e-02$  &  -1&-1&-1&$\phantom{-}4.437037e-01$  \\ 
%\lasthline
\hline	         	       
\hline
\end{tabular}					 
% \label{tabspec}
\end{table} 
   
All of the overlap integrals $I^k_{lmn}$ that appear in (\ref{eq:DR}) 
can be computed from the values in the tables using the relations 
(\ref{eq:H4}), (\ref{eq:H10}) and (\ref{eq:H11}).  There are only a 
finite number of these integrals that are non-zero, so they can be 
computed once and stored. 

%section{Calculation of $J_{mn} (k)$}

The second integral that is needed as input to equation 
(\ref{eq:DR}) is $J_{mn} (l)$.  
The first step to compute $J_{mn}(l)$ is to compute $\bar J_{mn} (l)$ 
defined in (\ref{eq:H9}).     
With a change of variables this integral can be rewritten as 
\[
\bar J_{mn} (k) = \int_{-\infty}^{\infty} dx' \int_{-\infty}^{\infty} 
dy' {\phi (x') \phi (y') \over 
k-m-n -x'- y'+i0^+}
\]
\beq
= \bar J_{k-m-n}  
\label{eq:I1}
\eeq
with
\beq
\bar{J}_{n} := \int_{-\infty}^\infty 
dx \int_{-\infty}^{\infty} 
dy {\phi (x) \phi (y) \over 
n -x- y+i0^+} .
\label{eq:I2}
\eeq

The support, $[0,2K-1]$, of the scaling function implies that in this
integral $x+y$ ranges from $0$ to $4K-2$.  This means that for $\vert
n \vert > (4K-2)$ the series
\beq
\bar J_n = { 1 \over n } \sum^{\infty}_{k=0} \int_{-\infty}^{\infty} dx 
\int_{-\infty}^{\infty} dy { 1 \over n^k} (x+y)^k \phi (x) \phi (y) 
\label{eq:I4}
\eeq
converges.  Using the binomial theorem we can express the integrals in
this series in terms of the known moments (\ref{eq:CU}) \cite{kessler03}
of the scaling function
\beq
\bar J_n = { 1 \over n } \sum^{\infty}_{m=0} \sum^{m}_{k=0} 
{ 1 \over n^m} {m! \over k!(m-k)! } 
\langle x^k \rangle \langle x^{m-k} \rangle
\label{eq:I5}
\eeq
This series converges rapidly for large $n$.  If $\bar{J}_n(N)$ is the 
approximation defined by summing the first $N$ terms 
of the series (\ref{eq:I5}) it follows that 
\beq
\vert \bar{J}_n - \bar{J}_n (N) \vert < \left ({(4K-2) \over 
\vert n \vert }\right )^{N+1} {(2K-1)^2 \over\vert  n-4K+2 \vert} \phi^2_{max},
\eeq
where $\phi_{max}$ is the maximum value $(< 1.5$ for $K=3)$ of the
scaling function.  For $\vert n \vert  \gg 4K-2$ this error can be made as
small as machine accuracy for modest values of $N$.

Thus for large $\vert n \vert $ the integrals $\bar{J}_n$ can be
computed efficiently and accurately by truncating the sum in
(\ref{eq:I5}). 
Integrals $\bar{J}_n$ for different values of $n$ are related by the
scaling equation (\ref{eq:H6}) which when used in (\ref{eq:I2}) gives
the homogeneous linear scaling equations for $\bar{J}_n$:
\beq
\bar J_{n} 
= \sum_{ll'}h_l h_{l'} \bar J_{2n-l-l'}. 
\label{eq:I3}
\eeq
Equation (\ref{eq:I3}) can be used to calculate $\bar{J}_n$
recursively using $\bar{J}_m$ for large $\vert m \vert $ as input.
This recursion can be used to step up in negative $n$ until $n=-1$ and
down in positive $n$ until $n=4K-1$.  This provides an efficient and
accurate method for calculating all of the $\bar{J}_n$ for $n<0$ and
$n>4K-2$.

The remaining values, $0 \leq n \leq 4K-2$, correspond to
cases where the denominator of the singular integral (\ref{eq:I2}) 
vanishes on the support of the integrand.

The scaling relations (\ref{eq:I3}) are still satisfied for these
values of $n$, giving $4K-1$ equations relating the unknown $\bar{J}_0
\cdots \bar{J}_{4K-2}$ to the known values of $\bar{J}_n$ for $n<0$
and $n>4K-2$.  Unlike the equations for $I_{lmn}$, these equations
cannot be linearly independent because they do not specify the
treatment of the singular integral.  One more equation is needed.
 
The desired equation can be derived by observing that the integral $\bar{J}_n$
can be expressed in terms of the autocorrelation function \cite{beylkin1}
\cite{beylkin2}\cite{beylkin3} 
$\Phi(x)$ of the scaling function as 
\beq
\bar J_{n} 
= - \int_{-\infty}^{\infty} {\Phi (y) \over y-n -i0^+ } dy
\label{eq:I5a}
\eeq
where  
\beq
\Phi(x) := \int^{\infty}_{-\infty} \phi(x-y) \phi (y) dy .
\label{eq:I6}
\eeq
It follows from the properties 
\beq
\int \phi (x) dx =1 \qquad 1 = \sum_n \phi (x+n) 
\label{eq:I7}
\eeq
of the scaling function that the autocorrelation function satisfies
\beq
\int \Phi (x) dx =1 \qquad 1 = \sum_n \Phi (x+n) 
\label{eq:I8}
\eeq
and has support on $[0 , 4K-2]$.  The autocorrelation function  
is plotted in Figure 3.

\begin{figure}
\begin{center}
\rotatebox{270}{\resizebox{2.9in}{!}{
\includegraphics{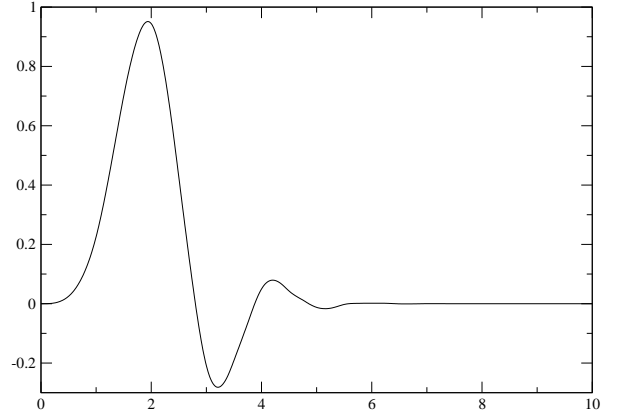}}
}
\end{center}
\label{Fig:3}
\caption{Daubechies' $K=3$ autocorrelation function.}
\end{figure}

Using (\ref{eq:I8}) in (\ref{eq:I6}) 
gives the additional linear constraint on the 
integrals $\bar{J}_n$:
\[
- i\pi =
- \int^m_{-m} { dx \over x - i 0^+ }=  
\]
\[
- \sum_n \int^m_{-m} dx {  \Phi(x+n) \over x - i0^+ }=
\]
\beq  
- \sum_n \int^{n+m}_{n-m} dx  { \Phi(x) \over x-n - i0 ^+ }  
\label{eq:I9}
\eeq
which holds for any $m$.  Replacing $-i\pi$ on the left side of 
equation (\ref{eq:I9}) by zero gives the principal value; by
$+i0^+$ gives the limit on the other side of the real line.

If $m > 4K-2 $  equation (\ref{eq:I9}) can be expressed as
\[
- i\pi  =
\sum_{n=-m}^{m} \bar{J}_n 
\]
\[
+\sum_{n=m+1}^{m+4K-3}\int_{n-m}^{4K-2} {\Phi (x) \over n-x +i0^+} dx
\]
\beq
+
\sum_{n=-m+1}^{-m+4K-3} \int_{0}^{n+m} {\Phi (x) \over n-x+i0^+ } dx .
\label{eq:I10}
\eeq
For $\vert n \vert >4K-2$ the boundary integrals 
\beq
\int_{n-m}^{4K-2} {\Phi (x) \over n-x+i0^+} dx =
\int_{n-m}^{\infty} {\Phi (x) \over n-x+i0^+ } dx 
\label{eq:I11}
\eeq
and
\beq
\int_{0}^{n+m} {\Phi (x) \over n-x+i0^+} dx =
\int_{-\infty}^{n+m} {\Phi (x) \over n-x+i0^+ } dx
\label{eq:I12}
\eeq
can be expanded in a convergent power series 
in terms of partial moments of the autocorrelation 
function 
\beq
\int_{n-m}^{4K-2} {\Phi (x) \over n-x+i0^+} dx =
{1 \over n} \sum_{k=0}^{\infty} {1 \over n^k} 
\int_{n-m}^{\infty} \Phi (x)x^k dx 
\label{eq:I13}
\eeq
and
\beq
\int_{0}^{n+m} {\Phi (x) \over n-x+i0^)} dx =
{1 \over n} \sum_{k=0}^{\infty} {1 \over n^k} 
\int_{-\infty}^{n-m} \Phi (x)x^k dx 
\label{eq:I14}
\eeq
The error after truncating the series in (\ref{eq:I13}) or (\ref{eq:I14})
after $N$ terms is bounded by 
\beq
\left ( {4K-2 \over n}\right )^{N+1} {4K-2 \over n -4K+2}\Phi_{max}  
\label{eq:I141}
\eeq
where $\Phi_{max} <1$ is the maximum value of the autocorrelation
function.  It is easy to compute these quantities to machine accuracy.
The partial moments of the autocorrelation function in
(\ref{eq:I13}) and (\ref{eq:I14}), which are needed
as input to (\ref{eq:I10}) can be computed analytically.  This
calculation is discussed in the appendix.

Equations (\ref{eq:I3}) and (\ref{eq:I10}) can be solved for the
$\bar{J}_n$ for $0 \leq n \leq 4K-2$ in terms of left side of
(\ref{eq:I10}), the partial moments of the autocorrelation function,
and the integrals for $\vert n\vert > 4K-2$.  The results of these
calculations are the nine complex numbers in Table 5.  This gives
values of $\bar{J}_m$ for all $m$.  The solution for principal value
is given by the real values in table 5, while thc conjugate of the
values in table gives the singular integral approaching the real line
from the other side.

\begin{table} % [hbt]
{\bf Table 5 - $\bar{J}_n$} \\[1.0ex]
\begin{tabular}{ c c }
%\firsthline
\hline
\hline
m & J  \\ 
\hline					    		      
0   &$-6.400535e-01 +i 0.000000e+00$   \\  
1   &$-1.570288e+00 -i 7.088321e-01$   \\   
2   &$\phantom{-}6.615596e-01 -i 2.966393e+00$   \\    
3   &$\phantom{-}1.719674e+00 +i 6.560028e-01$   \\    
4   &$\phantom{-}6.721642e-02 -i 1.554882e-01$   \\     
5   &$\phantom{-}3.595012e-01 +i 3.858342e-02$   \\    
6   &$\phantom{-}2.261977e-01 -i 5.023224e-03$   \\   
7   &$\phantom{-}1.853414e-01 -i 4.383938e-04$   \\    
8   &$\phantom{-}1.569998e-01 -i 3.586652e-06$   \\    
9   &$\phantom{-}1.357112e-01 -i 4.424016e-09$   \\    
10  &$\phantom{-}1.195044e-01 +i 0.000000e+00$   \\    
%\lasthline
\hline		       
\hline
\end{tabular}					 
% \label{tabspec}
\end{table}

The quantity that appears in the integral equation (\ref{eq:DR}) is 
$J_{mn}(k)$.  In order to evaluate this quantity first note
that it is symmetric in $m$ and $n$ so we can assume $m \geq n$.  
For $n \geq 0$ we can express 
$J_{mn}(k)$ in terms of $\bar{J}_{n}$;  
\beq
J_{mn}(k)=\bar{J}_{k-m-n} .
\label{eq:I15}
\eeq
When either $m$ or $n$ is less than $-2K+2$ then
\beq
J_{mn}(k)=0
\label{eq:I16}
\eeq 
The non-trivial values of $J_{mn}(k)$  correspond to $m$ non-negative 
and $-2K+2 \leq n \leq -1$, and both $-2K+2 \leq n,m \leq -1$.  This still
includes an infinite number of integrals because $k$ can take on any
value.

We discuss the treatment of $m$ non-negative and $-2K+2 \leq m \leq
-1$ separately.  When $m$ is non-negative the integral becomes
\beq
J_{mn} (k) = J_n (k-m)
\label{eq:I17}
\eeq 
where 
\beq J_n (m) := 
\int_{0}^{\infty} dx \int_{-\infty}^{\infty}
dy {\phi (x-n) \phi (y) \over m-x-y+i0^+}.
\label{eq:I18}
\eeq
Because in this case $n$ is within $2K-2$ of zero, for large $\vert m \vert $ 
this can be computed in terms of moments (\ref{eq:CU}) and partial 
moments (\ref{eq:CV}) of the scaling function using the series method:
\beq 
J_n (m) =
\sum_l {1 \over (m-n)^{l+1}} \sum_{k=0}^l \langle x^k \rangle_{[-n,2K-1]}
\langle x^{l-k} \rangle.
\label{eq:I19}
\eeq
The error made by keeping $N$ terms in the $l$ sum is bounded by
\beq
\left ( {2K-1 \over m-n}\right )^{N+1} 
{(2K-1)^2 \over m-n-2K+1}(\phi_{max})^2  
\eeq
which can be made as small as desired by choosing a large enough $m$. 

Using (\ref{eq:H6}) in (\ref{eq:I18}) gives
\beq
J_n (m) = \sum_{l=0}^{2K-1} \sum_{l'=0}^{2K-1} h_l h_{l'} J_{2n+l} (2m-l') 
\label{eq:I20}
\eeq
relations among these integrals for different values of $m$ and $n$.
Equation (\ref{eq:I20}) can be used to recursively step down from large
values of $\vert m \vert $ to $m=-1$ from below and $m=2K$ from above. 
Some terms in this recursion will be complex because they involve 
the integrals $\bar{J}_n$ for $1 \leq n \leq 4K-3$.

The values of $J_n(m)$ that cannot be computed directly from the
moments or by using the recursion correspond $-2K+2 \leq n \leq -1$
and $0 \leq m \leq 2K-1$.  These can be computed by treating
(\ref{eq:I20}) as a system of linear equations for the unknown
$J_n(m)$s.  This works because the terms in (\ref{eq:I20}) include
some of the previously computed integrals.  The results of this
calculation are shown in Table 6.
 
\begin{table} % [hbt]
{\bf Table 6 - $J_{m}(n)$} \\[1.0ex]
\begin{tabular}{ c c c }
%\firsthline
\hline
\hline
m & n & $J$ \\ 
\hline					    		      
-4 &  0 &$-5.270534e-05+i0.000000e-00$ \\         
-4 &  1 &$\phantom{-}1.328492e-04 -i1.360295e-03$ \\   
-4 &  2 &$\phantom{-}4.129155e-04 +i3.906821e-04$ \\   
-4 &  3 &$-9.487772e-05 -i9.892161e-05$ \\   
-4 &  4 &$\phantom{-}3.178448e-06-i1.800341e-06$  \\    
-4 &  5 &$\phantom{-}1.039698e-07-i6.117942e-09$ \\   
%-4 &  6 &$\phantom{-}6.581840e-05-i0.000000e-00$ \\   
%-4 &  7 &$\phantom{-}5.517594e-05-i0.000000e-00$ \\                       
-3 &  0 &$-4.625359e-03+i0.000000e-00$\\  
-3 &  1 &$-2.620525e-03-i5.640861e-02$ \\ 
-3 &  2 &$\phantom{-}2.569291e-02+i1.368533e-02$ \\ 
-3 &  3 &$-2.507103e-04-i2.505784e-03$ \\ 
-3 &  4 &$-1.189214e-03-i3.325919e-04$ \\ 
-3 &  5 &$\phantom{-}8.403857e-05+i4.945118e-06$  \\
%-3 &  6 &$2.802762e-03+i0.000000e-00$  \\  
%-3 &  7 &$2.349877e-03+i0.000000e-00$ \\ 
-2 &  0 &$\phantom{-}5.719523e-02+i0.000000e-00$ \\ 
-2 &  1 &$\phantom{-}8.006177e-02+i3.624922e-01$ \\	       
-2 &  2 &$-2.606313e-01-i7.778517e-02$ \\
-2 &  3 &$\phantom{-}3.145142e-02+i2.488590e-02$\\				 
-2 &  4 &$-1.188904e-02-i5.439550e-03$\\
-2 &  5 &$-9.599595e-03-i4.332178e-04$\\
%-2 &  6 &$-1.864109e-02-i3.603587e-06$ \\
%-2 &  7 &$-1.564854e-02-i4.424016e-09$ \\
-1 &  0 &$-2.730099e-01+i0.000000e-00$ \\ 
-1 &  1 &$-4.302964e-01-i1.483101e+00$\\ 
-1 &  2 &$\phantom{-}1.269047e+00+i2.936519e-01$\\ 
-1 &  3 &$-2.565108e-01-i9.893945e-02$\\ 
-1 &  4 &$\phantom{-}1.173431e-01+i4.010587e-02$\\ 
-1 &  5 &$\phantom{-}4.476128e-02-i5.054775e-03$\\ 
%-1 &  6 &$\phantom{-}7.609316e-02-i4.382921e-04$\\ 
%-1 &  7 &$\phantom{-}6.457402e-02-i3.586652e-06$\\ 			
%\lasthline
\hline
\hline		
\end{tabular}	
% \label{tabspec}	
\end{table}	

What remains are the $J_{mn}(k)$ when both $m$ and $n$ fall between
$-2K+2$ and $-1$.  In this case when $\vert k \vert $ is large,
$J_{mn}(k)$ can be expressed in the form of a convergent power series
in terms of moments and partial moments of the scaling function.	
The scaling equations can be used to step up or down in $k$ until
$k$ is between $0$ and $2K-2$; in addition they can also be used to 
solve for cases when
\beq
m,n,k \in [-2K+2,-1] \times [-2K+2,-1] \times [0,2K-1].
\label{eq:I21}
\eeq
In a normal application the value $k$ represents the on-shell
energy.   It will be large if there are a lot of basis functions 
with support on either side of the on-shell point.  While $J_{mn}(k)$ can
be calculated at the points (\ref{eq:I21}) from the known values 
using the scaling equation (\ref{eq:I20}), these points 
do not arise in most applications.

This completes the computation of the singular integrals that appear in 
equation (\ref{eq:DR}).

\section{Conclusion} 

In this paper we introduced a method for applying wavelet numerical
analysis to solve the relativistic three-body problem.  The method
starts by making variable changes in the relativistic 
Faddeev-Lovelace equations so the moving scattering singularity has 
simple scaling properties.

The next step is to project the equation in the transformed variables
on a finite resolution subspace of the three-body Hilbert space.  The
matrix representation of the integral equation in this approximation
space is easily computed by evaluating the driving terms and smooth
part of the kernel at the one-point quadrature points.  Additional
integrals involving the singular part of the kernel over the basis
functions are needed to compute the full kernel.  These integrals can
be calculated either exactly or with precisely controlled errors using
the scaling equation (\ref{eq:CA}) and normalization condition
(\ref{eq:CB}).  Method for computing all of the required integrals are
discussed in detail in section IV and the appendix.  Explicit values
of most of the needed integrals are computed and appear in Tables 3-6.  
The physics input 
is in the driving term and smooth part of the kernel.  The integrals 
over the singular part of the kernel are independent of the 
dynamics.  They can be computed from the scaling coefficients
by solving a small system of linear equations.
 
The resulting system of equations in the high resolution basis, while
easy to compute, is large.  The wavelet transform is used to transform
matrix elements in the high-resolution scaling basis to matrix
elements in a basis consisting of low resolution scaling functions and
wavelet basis functions with resolutions that fall between the
high-resolution and low resolution basis.  The transformation to this
new basis take $O(N)$ steps, which is faster than a fast Fourier
transform.  In the new basis the kernel can naturally be expressed as
the sum of a sparse matrix and a small matrix.  The key approximation
is to replace the small part of the kernel by zero.  The size of the
error made in this approximation can be controlled by changing the
threshold size for discarding matrix elements.

The sparse matrix can be solved using sparse matrix techniques.  In
reference \cite{kessler04} this was done by first conditioning the
matrix and then using the complex bi-conjugate gradient method.
The resulting solution can be transformed back to the scaling basis 
using the inverse wavelet transform.  The approximation is 
improved if the resulting solution is substituted back in the 
original equation \cite{Sloan}.  The step has the added benefit that
the basis functions never have to be calculated.
				   
The key result that was needed to calculate integrals associated with
the moving singularities is the observation by Beylkin \cite{beylkin1}
\cite{beylkin2}\cite{beylkin3} and collaborators that integrals of
scaling functions over moving singularities can be expressed as
integrals of the autocorrelation function of the scaling function over
a fixed singularity.  This leads to a practical and stable method for
computing the integrals.  The methods do not require subtractions or
careful choices of quadrature points; for the Daubechies' $K=3$ basis
they are reduced to solving a system of eleven linear equations.  The
required properties of the autocorrelation function are derived in the
appendix.

The research in references \cite{kessler03} and \cite{kessler04}
demonstrated that the wavelet method led to sparse matrix
approximations resulting negligible errors.  The structure of the
kernel in the relativistic three-body case indicates that the wavelet
method will lead to accurate sparse matrix approximations to the
relativistic Faddeev-Lovelace equations.

The increase in efficiency in this method is due to the saving in
computational effort in going from solving a large dense set of linear
equations to an approximately equivalent equations with a sparse
matrix.  The wavelet method will lead to a significant savings in
computational effort for a large system.

Our conclusion is that wavelet numerical analysis can be used to 
accurately approximate the relativistic Faddeev Lovelace equations 
by a linear system of equations with a sparse kernel matrix. 

\begin{acknowledgments}
This work supported in part by the Office of Science of the U.S. 
Department of Energy, under contract DE-FG02-86ER40286.  The authors 
acknowledge discussions with Fritz Keinert and Gerald Payne 
that contributed materially to this work.
\end{acknowledgments}

\appendix
%\section{Appendix}
\section{} 
%{Autocorrelation function of the scaling function}

The autocorrelation function of the scaling function is defined by 
\beq
\Phi(x) = \int^{\infty}_{-\infty} \phi(x-y) \phi (y) dy .
\label{eq:K1}
\eeq
Because of the support of the scaling function is $[0,2K-1]$, the 
autocorrelation function has support $[0,4K-2]$.

Using the scaling equation (\ref{eq:H6}) for the scaling function in
the definition (\ref{eq:K1}) of the autocorrelation function leads to 
the scaling
equation for the autocorrelation function:
\beq
\Phi (x) = 
 \sum_l \sum_{l'} h_l h_{l'} \Phi (2x -l'-l).
\label{eq:K2}
\eeq
If we define 
\beq
a_l = {1 \over \sqrt{2}} \sum_{l'=0}^{min(l,2K-1)} h_{l-l'} h_{l'} 
\qquad 0 \leq l \leq 4K-2
\label{eq:K3}
\eeq
equation (\ref{eq:K2}) can be put in the same form as (\ref{eq:CA}):
\beq
D \Phi (x) = 
\sum_{l=0}^{4K-2}  a_l T^l \Phi (x).
\label{eq:K4}
\eeq
The scaling coefficients $a_k$ for the Daubechies' $K=3$ 
autocorrelation function are given in Table 7. 
\begin{center}
\begin{table} % [hbt]
{\bf Table 7: Daubechies' $K=3$ Autocorrelation Scaling Coefficients } \\[1.0ex]
\begin{tabular}{ c c }
\hline
\hline
% $h_l$ & $K=3$   \\
%\hline					      		      
$a_0$ &$\phantom{-}7.825529e-02$ \\
$a_1$ &$\phantom{-}3.796160e-01$ \\
$a_2$ &$\phantom{-}6.767361e-01$  \\
$a_3$ &$\phantom{-}4.612557e-01$   \\
$a_4$ &$-4.471656e-02$ \\
$a_5$ &$-1.687321e-01$  \\
$a_6$ &$-2.481571e-03$  \\
$a_7$ &$\phantom{-}3.922363e-02$   \\
$a_8$ &$-1.563883e-03$  \\
$a_9$ &$-4.256471e-03$  \\
$a_{10}$ &$\phantom{-}8.774429e-04$ \\
\hline
\hline
\end{tabular}
% \label{tabspec}
\end{table}
\end{center}
\begin{center}
\begin{table} % [hbt]
{\bf Table 8: Daubechies' $K=3$ Autocorrelation Moments } \\[1.0ex]
\begin{tabular}{ c c }
\hline
\hline
% $\langle x^n \rangle_{\Phi}$ & $K=3$   \\  
% \hline					     		      
$\langle x^0 \rangle_{\Phi}$ &$1.000000e+00$ \\
$\langle x^1 \rangle_{\Phi}$ &$1.634802e+00$ \\
$\langle x^2 \rangle_{\Phi}$ &$2.672579e+00$\\
$\langle x^3 \rangle_{\Phi}$ &$4.167773e+00$\\
$\langle x^4 \rangle_{\Phi}$ &$5.825913e+00$\\
$\langle x^5 \rangle_{\Phi}$ &$6.817542e+00$\\
$\langle x^6 \rangle_{\Phi}$ &$8.807917e+00$\\
$\langle x^7 \rangle_{\Phi}$ &$4.055470e+01$\\
$\langle x^8 \rangle_{\Phi}$ &$2.899550e+02$\\
$\langle x^9 \rangle_{\Phi}$ &$1.695851e+03$\\
$\langle x^{10} \rangle_{\Phi}$ &$8.321402e+03$\\
\hline
\hline
\end{tabular}
% \label{tabspec}
\end{table}
\end{center}
The normalization condition (\ref{eq:CA}) of the scaling function can be used
in the definition (\ref{eq:K1}) of the autocorrelation function 
to derive the normalization condition:
\beq
\int \Phi (x) dx =1 .
\label{eq:K11}
\eeq

The scaling equation (\ref{eq:K2}) and the normalization 
condition (\ref{eq:K11}) can be used calculate moments and 
partial moments of the autocorrelation function.
To calculate the moments of 
the autocorrelation function use
\beq
\langle x^k \rangle_{\Phi} = \int \Phi (x) x^k dx,
\eeq
\beq
\langle x^k \rangle_{\Phi} = {1 \over 2^{k+1/2}} \sum_l a_l  
\sum_{n=0}^k {k! \over n! (k-n)!} l^{n-k}  \langle x^n \rangle_{\Phi} .
\eeq
Moving the $n=k$ term to the left side of the equation gives
recursion relation
\beq
\langle x^k \rangle_{\Phi} := 
{1 \over 2^k - 1}
{1 \over \sqrt{2} }\sum_l a_l  
\sum_{n=1}^k {k! \over n! (k-n)!} l^{n}  \langle x^{k-n} \rangle_{\Phi} .
\eeq
The recursion is started using the the normalization condition
(\ref{eq:K11}).  The lowest moments are tabulated in Table 8.

The partial moments of the autocorrelation function satisfy the 
scaling equation 
\[
\langle x^k \rangle_{\Phi, [m,\infty]}  := \int_m^{\infty}  \Phi (x) x^k dx =
\]
\beq
\sum_l a_l \sum_{n=0}^k {\sqrt{2} \over 2^{k+1}} 
{k! \over n! (k-n)!} l^m
\langle x^{k-m} \rangle_{\Phi,[2m-l,\infty]}.
\eeq
When $m \geq 4K-2$ these partial moments become ordinary moments while when 
$m\leq 0$ they vanish. This gives us a linear system for partial moments
in terms of the full moments and lower partial moments.  These
equations can be solved recursively. Partial moments 
corresponding to more general intervals can be computed by subtraction:
\beq
\langle x^{k} \rangle_{\Phi,[m,n]}=
\langle x^{k} \rangle_{\Phi,[m,\infty]} - \langle x^{k} \rangle_{\Phi,[n,\infty]} .
\eeq

\end{document}